\documentclass[reprint, superscriptaddress,  aps]{revtex4-2}
\usepackage{amssymb}
\usepackage{amsmath}
\usepackage{graphics}
\usepackage{graphicx}
\usepackage{subfigure}
\usepackage{dcolumn}
\usepackage{bm}
\usepackage{hyperref}
\usepackage{color}
\usepackage{epstopdf}

\definecolor{Gray}{gray}{0.1}
\DeclareMathAlphabet\mathbfcal{OMS}{cmsy}{b}{n}

\begin{document}

\title{High-harmonic spectroscopy of mobility edges in one-dimensional
quasicrystals}
\author{H. K. Avetissian}
\affiliation{Centre of Strong Fields Physics at Research Institute of Physics, Yerevan State University,
Yerevan 0025, Armenia}
\author{B. R. Avchyan}
\affiliation{Centre of Strong Fields Physics at Research Institute of Physics, Yerevan State University,
Yerevan 0025, Armenia}
\author{A. Brown}
\affiliation{Department of Chemistry, University of Alberta, Edmonton, AB, T6G 2G2, Canada}
\author{G. F. Mkrtchian}
\thanks{mkrtchian@ysu.am}
\affiliation{Centre of Strong Fields Physics at Research Institute of Physics, Yerevan State University,
Yerevan 0025, Armenia}

\begin{abstract}
Quasicrystals occupy a unique position between periodic and disordered
systems, where localization phenomena such as Anderson transitions and
mobility edges can emerge even in the absence of disorder. This distinctive
behavior motivates the development of robust, all-optical diagnostic tools
capable of probing the structural, topological, and dynamical properties of
such systems. In this work, focusing on generalized Aubry-Andr\'{e}-Harper
models and on an incommensurate potential in the continuum limit, we demonstrate that high-harmonic generation phenomenon serves as a
powerful probe of localization transitions and mobility edges in
quasicrystals. We introduce a new parameter--dipole mobility--which captures
the impact of intraband dipole transitions and enables classification of
nonlinear optical regimes, where excitation and high-harmonic generation
yield can differ by orders of magnitude. We show that the cutoff frequency
of harmonics is strongly influenced by the position of the mobility edge,
providing a robust and experimentally accessible signature of localization
transitions in quasicrystals.
\end{abstract}

\maketitle

\section{Introduction}

Quasicrystals constitute an intermediate case between periodic and
disordered systems. In disordered systems it is possible Anderson phase
transition, when single particle wave functions shift from extended to
exponentially localized beyond a critical disorder strength \cite%
{anderson1958absence}. The scaling theory reveals \cite%
{abrahams1979scaling,evers2008anderson} that in one-dimensional (1D)
Anderson models, this phase transition occurs even at the exceedingly weak
disorders. As a result, the 1D Anderson localization becomes somewhat less
intriguing when viewed in the context of the physics behind disorder-driven
metal-insulator transitions. Conversely, in the case of three dimensions
(3D), a mobility edge (ME) emerges within the single-particle spectrum,
serving as a critical point that distinguishes extended eigenstates from
their localized counterparts \cite{mott1987mobility}. In quasicrystals, the
metal-to-insulator transition manifests even in 1D systems, as exemplified
by the well-known Aubry-Andr\'{e}-Harper (AAH) model \cite%
{aubry1980analyticity,harper1955single}. This lattice model with
incommensurate on-site modulations plays a central role in the study of
Anderson localization in quasicrystals \cite%
{ostlund1983one,kohmoto1983metal,sarma1988mobility,sarma1990localization,biddle2009localization}%
. However, the self-dual symmetry inherent to the AAH model prevents the
observation of MEs. To introduce MEs, this self-duality must be disrupted.
Extensive research has shown that mobility edges can emerge through
mechanisms such as long-range hoppings \cite%
{biddle2010predicted,biddle2011localization,liu2021exact}, power-law decay
of hopping amplitudes \cite%
{PhysRevLett.123.025301,PhysRevB.103.075124,PhysRevB.83.075105,PhysRevB.100.174201,peng2023power}%
, and alterations to the incommensurate potential \cite%
{ganeshan2015nearest,li2020mobility,PhysRevLett.122.237601,PhysRevResearch.2.033052}%
, on the basis of which different new models were developed --Generalized
Aubry-Andr\'{e}-Harper (GAAH) models--to describe the MEs. Recently, the
study of quasiperiodic systems has garnered significant attention, driven by
hopeful advancements in experimental realizations of 1D AAH model \cite%
{roati2008anderson,lahini2009observation,kraus2012topological,lohse2016thouless,luschen2018single,an2021interactions,wang2022observation}%
. Consequently, the GAAH models have opened new possibilities in condensed
matter physics, offering insights into challenging problems \cite%
{PhysRevA.98.013635,PhysRevB.90.054303,PhysRevB.101.064203,PhysRevB.96.085119,PhysRevA.105.063327,huang2023incommensurate,Bordia_2017}
inherent to 3D systems, using a simpler model.

GAAH models display diverse physical phenomena, featuring adjustable energy
spectra and the simultaneous presence of extended, critical, and localized
single-particle states. This complexity highlights the need for advanced
diagnostic tools that can uncover the structural, topological, and dynamical
characteristics of these systems. Ideally, such tools should be versatile
and effective across various models and dimensions, enabling deeper
understanding of localization phenomena in realistic three-dimensional
settings.

High-harmonic spectroscopy \cite{peng2019attosecond}, based on the
high-harmonic generation (HHG) multiphoton phenomenon \cite{corkum1993plasma}%
, has emerged as a powerful technique for probing ultrafast electronic
dynamics. Initially studied in atomic and molecular gases, HHG has now been
extended to crystals \cite{ghimire2019high,ghimire2014strong}, quasicrystals 
\cite{liu2021multichannel,liu2024distinguishing} and liquids \cite%
{zeng2020impact,xia2022theoretical}, unlocking new possibilities in extreme
solid-state photonics \cite%
{vampa2017merge,ghimire2019high,li2020attosecond,brabec2022high}. HHG has
been employed to probe electronic structure \cite%
{vampa2015all,luu2015extreme,lakhotia2020laser}, band topology \cite%
{kelardeh2017graphene,bauer2018high,silva2019topological,chacon2020circular,avetissian2020high,schmid2021tunable,bai2021high,pattanayak2022role,avetissian2022high,avetissian2025berry}%
, electron correlations \cite%
{vampa2015linking,tancogne2017impact,silva2018high,avetissian2020many,murakami2021high,neufeld2021light}%
, and localization phenomena \cite%
{orlando2022ellipticity,zeng2023periodic,avetissian2023disorder,avetissian2024intense,dziurawiec2024}%
.

In the current work, we focus on two specific realizations of the GAAH
model: the Biddle-Das Sarma (BD) model \cite{biddle2010predicted} and the
Ganeshan-Pixley-Das Sarma (GPD) model \cite{ganeshan2015nearest}. Besides,
we also investigate HHG in a one-dimensional bichromatic incommensurate
potential in the continuum limit, beyond the tight-binding approximation. We
demonstrate that HHG is highly sensitive to the presence of MEs, providing a
new method for their detection. A related investigation for the GPD model
was recently presented in Ref. \cite{dziurawiec2024}, where the dependence
of the HHG yield on the filling factor was analyzed. In the mentioned study,
a detailed comparison of below-band-gap high-harmonic generation spectra
between the GPD and conventional AAH models identified distinct spectral
features attributable to the emergence of MEs within the single-particle
spectrum. While our work builds on similar models, the core objective and
methodology of our study differ significantly from Ref. \cite{dziurawiec2024}%
. The cited work focuses on comparing HHG yields between AAH and GAAH models
as a function of filling factor using a system-dependent ratio. In essence,
to detect a mobility edge, one must compute HHG spectra for both systems,
while varying the Fermi energy--a process that is not only model-dependent
but also experimentally challenging. In the current investigation, we pursue
a different approach: we fix the filling and analyze the physical origin of
the HHG cutoff, showing that the cutoff frequency exhibits a robust and
universal dependence on the ME position. This allows the presence of a
mobility edge to be inferred from a single HHG spectrum, making the method
experimentally much simpler. Additionally, we extend the analysis beyond the
GPD model by including the BD model with long-range hopping and the
continuum model with an incommensurate potential--an important cases not
addressed in Ref. \cite{dziurawiec2024}.

To capture the HHG cutoff behavior, we introduce a new order parameter --
dipole mobility, which incorporates information about intraband dipole
transitions. Unlike conventional localization measures (e.g., inverse
participation ratio (IPR) or Shannon entropy), the dipole mobility directly
relates to nonlinear optical response and provides deeper insight into
localization transitions. This diagnostic is intuitive, model-independent,
and applicable to more realistic systems of higher dimensions.

This paper is organized as follows. In Section II, the model and the basic
equations are formulated. In Section III, we represent the main results.
Conclusions are given in Section IV. In Appendix A we consider various
excitation regimes in a system with two energy bands.

\section{The model and the basic equations}

\subsection{The tight-binding models}

We begin by outlining the models and theoretical framework. We consider a
family of 1D quasicrystals described within the tight-binding model, with
the Hamiltonian given by:%
\begin{equation}
\widehat{H}_{\mathrm{TB}}=\sum_{i\neq j}t_{ij}c_{i}^{\dagger
}c_{j}+\sum_{i=1}^{N}\Lambda \left( 2\pi \sigma i\right) c_{i}^{\dagger
}c_{i}.  \label{h0}
\end{equation}%
Here, $c_{i}^{\dagger }$\ ($c_{i}$) represents the creation (annihilation)
operator for a fermion at site $i$, and $t_{ij}$ is the hopping integral
between the sites $i$ and $j$. In Eq. (\ref{h0}), the on-site energy $%
\Lambda \left( \zeta \right) $ is a periodic function: $\Lambda \left( \zeta
\right) =\Lambda \left( \zeta +2\pi \right) $, but for an irrational $\sigma 
$, it is incommensurate with the underlying lattice. The chain consists of $%
N $ sites with open boundary conditions, where $i=1$ and $i=N$ represent the
edge sites. Throughout this work, we set $N=200$ and $\sigma =(\sqrt{5}-1)/2$%
, though any other irrational value of $\sigma $ is equally valid.

The first model we consider is the AAH model \cite%
{aubry1980analyticity,harper1955single}, characterized by an on-site
potential $\Lambda \left( \zeta \right) =2V\cos \zeta $ and nearest-neighbor
hopping $t_{ij}=J(\delta _{ji+1}+\delta _{ij+1})$, where $\delta _{ij}$ is
the Kronecker delta. Without loss of generality we take $V>0$ and $J>0$. In
the AAH model, when $V>J$, the single-particle electronic wavefunctions
become localized, similar to the Anderson model. Conversely, for $V<J$, the
wavefunctions are delocalized. The $V=J$ marks a self-dual condition where
the real-space and momentum-space representations are identical. The
breaking duality of the AAH model leads to an energy-dependent self-duality
relation, giving rise to MEs. For instance, introducing a second
incommensurate potential or including next-nearest-neighbor hopping results
in a ME. Among the several models, several allow for analytical solutions to
the ME problem. While our findings apply to a broad class of models, we
focus on the BD and GPD models. These models provide valuable insights into
localization and MEs in 1D quasicrystals. The BD model extends the AAH model
by introducing long-range hopping. The hopping amplitude decays
exponentially with distance as $t_{ij}=J\exp (-p\left\vert i-j\right\vert
+p) $, \textrm{where $p>0$ is the decay factor}. The GPD model modifies the
quasiperiodic potential in the AAH model as $\Lambda \left( \zeta \right)
=2V\cos \zeta /(1-b\cos \zeta )$, incorporating higher harmonics of the
incommensurate potential. In contrast to AAH model the GAAH models allow for
energy ($\mathcal{E}$) dependent duality points. The ME, separating the
localized and extended states for BD model, is given by the following
expression:%
\begin{equation}
\mathcal{E}=2V\cosh (p)-J\exp (p),  \label{BD}
\end{equation}%
while in GPD model the latter is defined by the expression: 
\begin{equation}
\mathcal{E}=\frac{2}{b}\left( J-V\right) .  \label{GBD}
\end{equation}%
Note that Eqs. (\ref{BD}) and (\ref{GBD}) come to that of the AAH model ($%
V=J $) for $p\rightarrow \infty $ and $b\rightarrow 0$, respectively.

We now couple the GAAH system with a strong linearly polarized incident
laser pulse. The light-matter interaction is described in the length-gauge
by the Hamiltonian 
\begin{equation}
\widehat{H}_{\mathrm{int}}=e\sum_{j}x_{j}E\left( t\right) c_{j}^{\dagger
}c_{j},  \label{int}
\end{equation}%
where $e$ is the fermion charge, $E\left( t\right) =f\left( t\right)
E_{0}\cos \omega t$ is the electric field strength, with the amplitude $%
E_{0} $, frequency $\omega $, and pulse envelope $f\left( t\right) =\sin
^{2}\left( \pi t/\mathcal{T}\right) $. The pulse duration $\mathcal{T}$ is
taken to be $15$ wave cycles: $\mathcal{T}=30\pi /\omega $.\ In Eq. (\ref%
{int}) $x_{j}=aj$ denotes the coordinate of the $j$-th lattice site, where $%
a $ is the lattice constant. To maintain generality and keep the discussion
model-independent, we adopt a system of local units: $a$ sets the unit of
length, the hopping amplitude $J$ defines the unit of energy, and the
elementary charge is set to $e=1$. Within this convention, the dimensionless
electric field amplitude $E_{0}$ corresponds to a physical electric field
strength of $eE_{0}a/J$.

From the Heisenberg equation one can obtain evolutionary equation for the
single-particle density matrix $\rho _{ij}=\left\langle c_{j}^{\dagger
}c_{i}\right\rangle $ : 
\begin{equation}
i\frac{\partial \rho _{ij}}{\partial t}=\sum_{k}\left( \tau _{kj}\rho
_{ik}-\tau _{ik}\rho _{kj}\right) +E\left( t\right) \left(
x_{i}-x_{j}\right) \rho _{ij},  \label{evEqs}
\end{equation}%
where $\tau _{ij}=-\Lambda \left( 2\pi \sigma i\right) \delta _{ij}-t_{ij}.$
To gain physical insight, it is useful to express the dynamics in the energy
eigenbasis. Performing the transformation $\rho _{ij}=\sum_{\mu \mu ^{\prime
}}\psi _{\mu ^{\prime }}^{\ast }\left( j\right) \varrho _{\mu \mu ^{\prime
}}\psi _{\mu }\left( i\right) $, where $\psi _{\mu }(i)$ are the
eigenfunctions corresponding to eigenenergies $\varepsilon _{\mu }$ of
Hamiltonian (\ref{h0}), and $\varrho _{\mu \mu ^{\prime }}$ is the density
matrix in the energy basis, Eq. (\ref{evEqs}) becomes: 
\begin{equation}
i\frac{\partial \varrho _{mn}}{\partial t}=\varepsilon _{mn}\varrho
_{mn}+E\left( t\right) \sum_{\mu }\left( \varrho _{\mu n}d_{m\mu }-\varrho
_{m\mu }d_{\mu n}\right) ,  \label{main}
\end{equation}%
where $d_{\mu ^{\prime }\mu }=\sum_{i}\psi _{\mu ^{\prime }}^{\ast }\left(
i\right) x_{i}\psi _{\mu }\left( i\right) $ is the transition dipole moment
and $\varepsilon _{mn}=\varepsilon _{m}-\varepsilon _{n}$. Relaxation
processes can be introduced in Eqs. (\ref{evEqs}) and (\ref{main})
phenomenologically via damping term, assuming that the system relaxes at a
rate $\gamma $\ to the equilibrium distribution. To compute the harmonic
spectrum, we evaluate the Fourier transform: $a\left( \Omega \right)
=\int_{-\infty }^{\infty }a\left( t\right) e^{i\Omega t}W\left( t\right) dt$%
, where $a\left( t\right) =\sum_{i}x_{i}d^{2}\rho _{ii}/dt^{2}$ is the
dipole acceleration and $W\left( t\right) $ is a window function that
reduces the overall background noise of the harmonic signal \cite%
{zhang2018generating}. We choose the pulse envelope $f\left( t\right) $ as a
window function. The emitted intensity at a given frequency is proportional
to the square of the Fourier amplitude, $\left\vert a\left( \Omega \right)
\right\vert ^{2}$. We perform the time integration of Eqs. (\ref{evEqs}) and
(\ref{main}) with the help of the eighth-order Runge-Kutta method. For the
solution of Eqs. (\ref{evEqs}) and (\ref{main}) we need initial density
matrix. To this end, we numerically diagonalize the tight-binding
Hamiltonian (\ref{h0}) and construct the initial density matrix $\rho
_{0ij}=\sum_{\mu =1}^{N/2}\psi _{\mu }^{\ast }(j)\psi _{\mu }(i)$ by filling
the lowest $N/2$ states (i.e., a half-filled system at the zero
temperature). 
\begin{figure*}[tbp]
\includegraphics[width=0.98\textwidth]{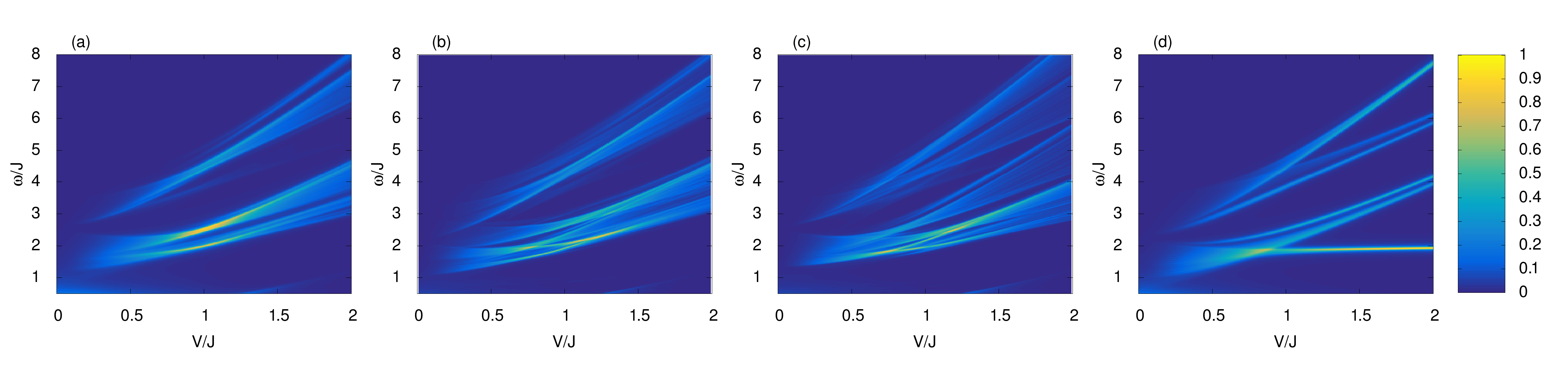}
\caption{Linear optical response of 1D quasicrystals as a function of the
amplitude of the on-site potential. The color scale represents the
normalized absorption coefficient: (a) AAH model, (b) BD model with $p=2.0$,
(c) GPD model with $b=0.275$, and (d) AAH model with commensurate modulation 
$\protect\sigma =2/5$. The relaxation rate is taken to be $\protect\gamma%
=0.01J$.}
\end{figure*}
Since we consider a half-filled system, by convention the occupied states
can be identified as the valence band, while the unoccupied states
correspond to the conduction band. It is important to note that, due to the
fractal nature of the quasiperiodic spectrum, these bands exhibit intricate
subband structures. Within this convention, we can meaningfully separate the
interband and intraband contributions to the HHG spectrum by decomposing the
dipole acceleration as follows: 
\begin{equation}
a\left( t\right) =a_{\mathrm{intra}}\left( t\right) +a_{\mathrm{inter}%
}\left( t\right) ,  \label{dec1}
\end{equation}
where 
\begin{equation}
a_{\mathrm{intra}}\left( t\right) =\sum_{\mu ,\mu ^{\prime }=N/2}^{N-1}\frac{%
d^{2}\rho _{\mu \mu ^{\prime }}\left( t\right) }{dt^{2}}d_{\mu ^{\prime }\mu
}+\sum_{\mu ,\mu =0}^{N/2-1}\frac{d^{2}\rho _{\mu \mu ^{\prime }}\left(
t\right) }{dt^{2}}d_{\mu ^{\prime }\mu },  \label{dec2}
\end{equation}%
is the intraband part involving transitions within the same group (valence
or conduction states) and

\begin{equation}
a_{\mathrm{inter}}\left( t\right) =\sum_{\mu ^{\prime }=N/2}^{N-1}\sum_{\mu
=0}^{N/2-1}\left( \frac{d^{2}\rho _{\mu \mu ^{\prime }}\left( t\right) }{%
dt^{2}}d_{\mu ^{\prime }\mu }\right) +\mathrm{c.c.},  \label{dec3}
\end{equation}%
is the interband part corresponding to coherence between occupied and
unoccupied states.

\subsection{Bichromatic incommensurate potential in the continuum limit}

We also investigate HHG in a one-dimensional bichromatic incommensurate
potential in the continuum limit, beyond the tight-binding approximation, by
solving the full time-dependent Schr\"{o}dinger (TDS) equation with two
potentials in the presence of a pump laser field. The Hamiltonian is

\begin{equation*}
\widehat{H}=\widehat{H}_{0}+exE(t),
\end{equation*}%
where 
\begin{equation}
\widehat{H}_{0}=-\frac{\hbar ^{2}}{2m}\frac{\partial ^{2}}{\partial x^{2}}
+V_{p}\cos \!\left( \frac{2\pi }{a}x\right) +V_{s}\cos \!\left( \frac{2\pi }{%
a}\sigma x\right)  \label{hc}
\end{equation}%
is the Hamiltonian of a 1D quasicrystal. Here $V_{p}$ and $V_{s}$ are the
amplitudes of the primary and secondary potentials, respectively. For
irrational $\sigma$, the system is aperiodic. Unlike the tight-binding
reduction, we solve the Schr\"{o}dinger equation exactly, treating the
primary and secondary potentials on equal footing.

The characteristic energy scale is the recoil energy of the primary lattice, 
$E_{r}=\hbar ^{2}\pi ^{2}/(2ma^{2})$. In the deep-lattice limit $V_{p}\gg
E_{r}$ and $V_{s}\ll V_{p}$, the physical properties of Eq.~(\ref{hc}) can
be mapped to the AAH model: the nearest-neighbor hopping $J$ is set by the
primary lattice depth $V_{p}$, while the disorder potential $V$ depends on
both $V_{s}$ and $V_{p}$ \cite{biddle2010predicted}. For shallow lattices,
however, a full continuum treatment is required.

For numerical solutions we employ local atomic units (l.a.u.): the length
unit $l=a\sqrt{2}/\pi $, the recoil energy $E_{r}$ as the energy unit, and $%
\hbar /E_{r}$ as the time unit. In these units, the TDS equation becomes 
\begin{equation*}
i\frac{\partial \Psi (x,t)}{\partial t}=\left[ -\frac{1}{2}\frac{\partial
^{2}}{\partial x^{2}}+\widetilde{V}_{p}\cos (2\sqrt{2}x)\right. 
\end{equation*}

\begin{equation}
\left. +\widetilde{V}_{s}\cos (2\sqrt{2}\sigma x)+x\widetilde{E}(t)\right]
\Psi (x,t),  \label{TDS}
\end{equation}%
where $\widetilde{V}_{p,s}=V_{p,s}/E_{r}$ and $\widetilde{E}(t)=elE(t)/E_{r}$%
. The TDS equation is solved with the OCTOPUS package \cite{andrade2015real}
in the independent-particle mode on a real-space grid, using a time step of $%
0.1$ l.a.u. The simulation domain has a length of $240$ l.a.u. with grid
spacing $0.1$ l.a.u. To avoid artificial reflections, a complex absorbing
potential is applied at the box boundaries with height $-0.85$ l.a.u. and
absorbing length $10$ l.a.u..

\section{Results}

Before presenting the main results on HHG in quasicrystals, we briefly
examine the linear optical response for completeness and to clearly
highlight the fundamentally different nature of the HHG signal compared to
the linear response. The linear response is determined by the initial
population, transition dipole matrix elements, and the joint density of
states. The conduction is defined via perturbative solution of Eq. (\ref%
{main}), as follows:%
\begin{equation*}
\sigma \left( \omega \right) =-2i\omega \sum_{m\leq
N/2}\sum_{n>N/2}\left\vert d_{nm}\right\vert ^{2}
\end{equation*}%
\begin{equation}
\times \left[ \frac{1}{\varepsilon _{n}-\varepsilon _{m}-\omega -i\gamma }+%
\frac{1}{\varepsilon _{n}-\varepsilon _{m}+\omega +i\gamma }\right] .
\label{sij}
\end{equation}
\begin{figure}[tbp]
\includegraphics[width=0.49\textwidth]{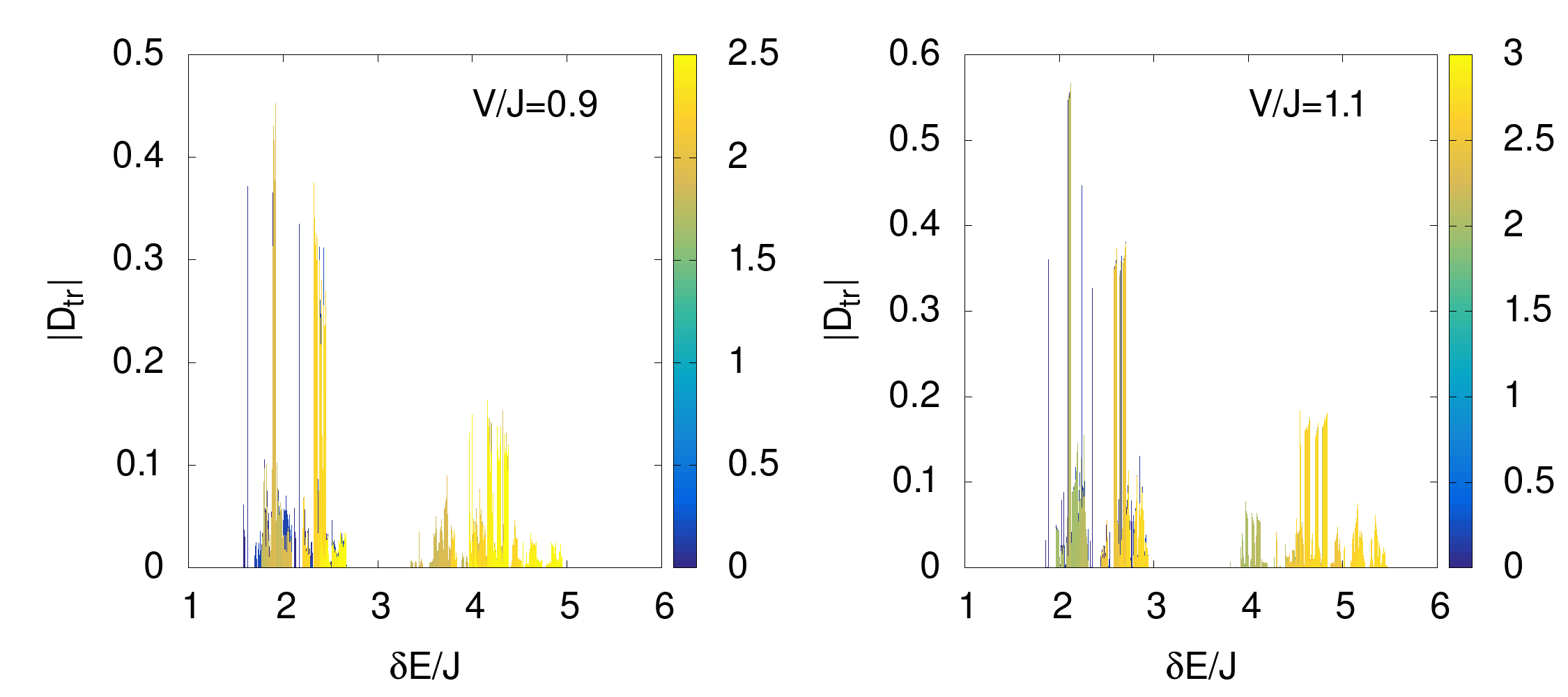}
\caption{Absolute values of the interband transition dipole matrix elements
for the AAH model as a function of transition energy. Left panel: Results
for delocalized states. Right panel: Results for localized states. The color
scale represents the energy ranges (in units of $J$) of the conduction
bands. }
\end{figure}
\begin{figure*}[tbp]
\includegraphics[width=0.98\textwidth]{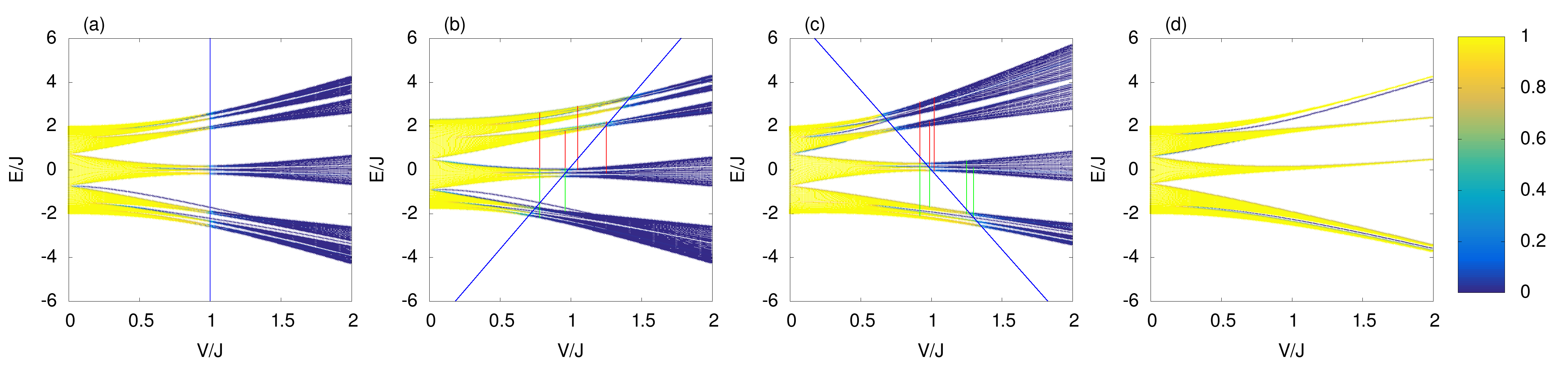}
\caption{Energy eigenvalues of 1D quasicrystals described by Hamiltonian (%
\protect\ref{h0}): (a) AAH model, (b) BD model with $p = 2.0$, (c) GPD model
with $b = 0.275$, and (d) AAH model with commensurate modulation $\protect%
\sigma = 2/5$. The shading of the energy levels represents the dipole
mobility $D_\protect\mu$ defined in Eq. (\protect\ref{DM}). Solid blue lines
denote the analytical boundaries between spatially localized and extended
states, given by $V/J = 1$ and Eqs. (\protect\ref{BD}) and (\protect\ref{GBD}%
). Vertical lines indicate transitions involving states near the MEs.}
\end{figure*}

As evident from Eq. (\ref{sij}), the conductivity is independent of
intraband transition dipole moments and therefore does not provide
information about localization or ME. To illustrate this, Fig. 1 presents
the real part of the optical conductivity, $\mathrm{Re}\sigma \left( \omega
\right) $, for the three models under consideration. Additionally, we
examine the AAH model with commensurate modulation, where no transition
occurs between the extended and localized states. As seen in Fig. 1, in all
cases, the linear optical response remains insensitive to the behavior of
single-particle wave functions. To further highlight this point, Fig. 2
displays the absolute values of the interband transition dipole matrix
elements for the AAH model as a function of transition energy. The results
demonstrate that the transition dipole moments remain of the same order of
magnitude for both localized and delocalized states. This further confirms
that linear response alone cannot capture the localization properties of
eigenstates in quasiperiodic systems.

In contrast to the linear regime of interaction, nonlinear effects,
particularly multiphoton excitations and HHG are highly sensitive to the
behavior of single-particle wave functions and average dipole moments \cite%
{brown2000rotating,avetissian2011coherent}. The localization properties of
the normalized eigenstates ($\sum_{i}\left\vert \psi _{\mu }\left( i\right)
\right\vert ^{2}=1$) are typically characterized using the inverse
participation ratio $IPR=\sum_{i}\left\vert \psi _{\mu }\left( i\right)
\right\vert ^{4}$. In the large $N$ limit it scales as $IPR\varpropto N^{-f}$%
, where $f$ represents the fractal dimension of the eigenstate: $f=0$ for
localized eigenstates, $f=1$ for extended eigenstates, and $0<f<1$ for
critical eigenstates. Another widely used localization measure is Shannon
entropy. However, both the IPR and Shannon entropy fail to capture
information about intraband dipole transitions. To address this limitation,
a new order parameter for the localization transition in quasiperiodic
systems is introduced: \textit{dipole mobility}, defined as 
\begin{equation}
D_{\mu }=\frac{2\sqrt{\pi }}{N}\left( \left\vert d_{\mu \mu +1}\right\vert
^{2}+\left\vert d_{\mu \mu -1}\right\vert ^{2}\right) ^{1/2}.  \label{DM}
\end{equation}%
This measure provides additional insight into the localization transition by
incorporating information about intraband transitions, which play a crucial
role in nonlinear optical response and electronic transport. In Fig. 3,
energy eigenvalues of 1D quasicrystals described by the Hamiltonian (\ref{h0}%
) are shown. The shading of the energy curves represents the dipole mobility
(\ref{DM}). The solid lines denote the analytical boundary between spatially
localized and spatially extended states. The spectral structure is fractal,
comprising a hierarchy of sub-bands that depend sensitively on the
incommensurate modulation parameter $\sigma $ and the amplitude of the
on-site potential $V$. Additionally, due to the use of open boundary
conditions, the system exhibits topologically protected edge states \cite%
{kraus2012topological}, which appear within spectral gaps and remain
localized regardless of the extended-to-localized transition of the bulk
states. As seen in Fig. 3, the dipole mobility closely resembles the
fundamental characteristics captured by the inverse participation ratio: $%
D_{\mu }=0$ for localized eigenstates, $D_{\mu }\simeq 1$ for extended
eigenstates. However, in addition to identifying localization properties,
dipole mobility enables us to classify the nonlinear interaction of
quasicrystals into three distinct regimes. These regimes are schematically
illustrated in Fig. 4, where excitation and subsequent HHG processes exhibit
probability differences spanning several orders of magnitude.

\begin{figure}[tbp]
\includegraphics[width=0.4\textwidth]{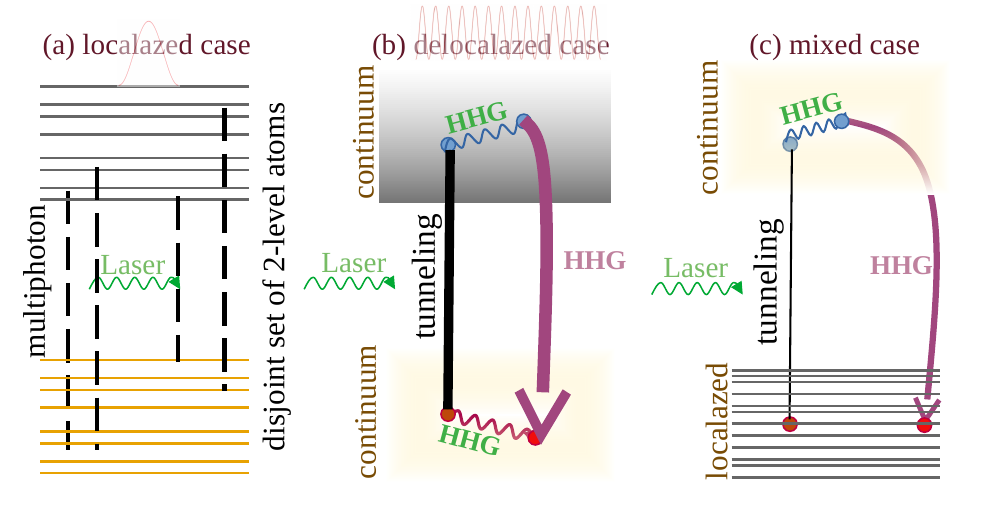}
\caption{Schematic illustration of three distinct interaction regimes in a
system with two energy bands, characterized by different localization
properties of single-particle wavefunctions, under a low-frequency laser
field.}
\end{figure}

We now turn to the nonlinear optical response of quasicrystals driven by a
strong laser field of frequency much smaller than the nearest-neighbor
hopping amplitude, i.e., $\omega \ll J$. Figure 4 depicts a schematic system
consisting of two energy bands subjected to such a low-frequency field.
Specifically, we assume that $\omega \ll \Delta $, where $\Delta $ is the
band gap. We use the term 'band' by analogy in the terminology of periodic
crystals to refer to energy regions separated by gaps much larger than the
photon energy of the driving laser, recognizing that the lack of
translational symmetry in quasicrystals makes conventional Bloch band
definitions imprecise. The field strength is chosen such that the system
lies in the multiphoton regime: $\Delta $ $\gg \left\vert d_{tr}\right\vert
E_{0}\gtrsim \omega $, meaning that the electric dipole interaction is
stronger than the energy of a single photon but still much smaller than the
band gap. In the case where states in both bands are fully localized ($%
D_{\mu }=0$), the system behaves effectively as a disjoint set of two-level
atoms [Fig. 4(a)]. Excitation in this case occurs via a multiphoton process 
\cite{shirley1965solution,duvall1988nonperturbative} with negligible
probability. Additionally, intraband transitions are absent, preventing
further excitation within the band. In the scope of the perturbative
framework, the details of which are provided in the Appendix A, the average
excitation rate can be estimated as $R_{pert}\propto \left( \left\vert
d_{tr}\right\vert E_{0}/\Delta \right) ^{\frac{\Delta }{\omega }}$. For $%
\Delta \gg \omega $ and $|d_{tr}|E_{0}/\Delta \ll 1$, this rate becomes
exponentially small, rendering excitation and subsequent HHG highly
inefficient.

In contrast, when the states within both bands are fully delocalized $D_{\mu
}\simeq 1$ (Fig. (4b)), the system operates in the tunneling excitation
regime, described by Keldysh theory \cite%
{keldysh1965ionization,vampa2014theoretical} in the thermodynamic limit $%
N\rightarrow \infty $. The presence of a continuum of states is essential
for tunneling and besides the wave electric field strength should be much
smaller than characteristic electric field strength $E_{a}\approx \Delta
/\left\vert d_{tr}\right\vert $ of the considered system. This corresponds
to the well-established three-step model of HHG. The first step involves
excitation via tunneling from the lower band to the upper band, creating an
electron-hole pair. The second step consists of further excitation and
de-excitation within the band, leading to intraband harmonic emission. The
final step is electron-hole recombination, resulting in interband harmonic
emission. The tunneling excitation rate $R_{tun}\propto e^{-\kappa
E_{a}/E_{0}}$, where $\kappa $ is a numerical constant of order unity,
attains significant values for fields $E_{a}/E_{0}\sim 10$, which is much
larger than the perturbative excitation rate. An intermediate scenario is
depicted in Fig. 4(c) when the states in one band are fully localized ($%
D_{\mu }=0$), while those in the other band are fully delocalized ($D_{\mu
}\simeq 1$). In this case, the system also operates in a tunneling
excitation regime, where transitions occur from isolated localized states to
a continuum of delocalized states, resembling atomic HHG \cite%
{lewenstein1994theory}. However, in this case, the excitation and
recombination amplitudes are reduced compared to the fully delocalized
scenario, as the number of available excitation channels is lower. These
regimes are considered in the Appendix A using a simple continuum model.

\begin{figure}[tbp]
\includegraphics[width=0.49\textwidth]{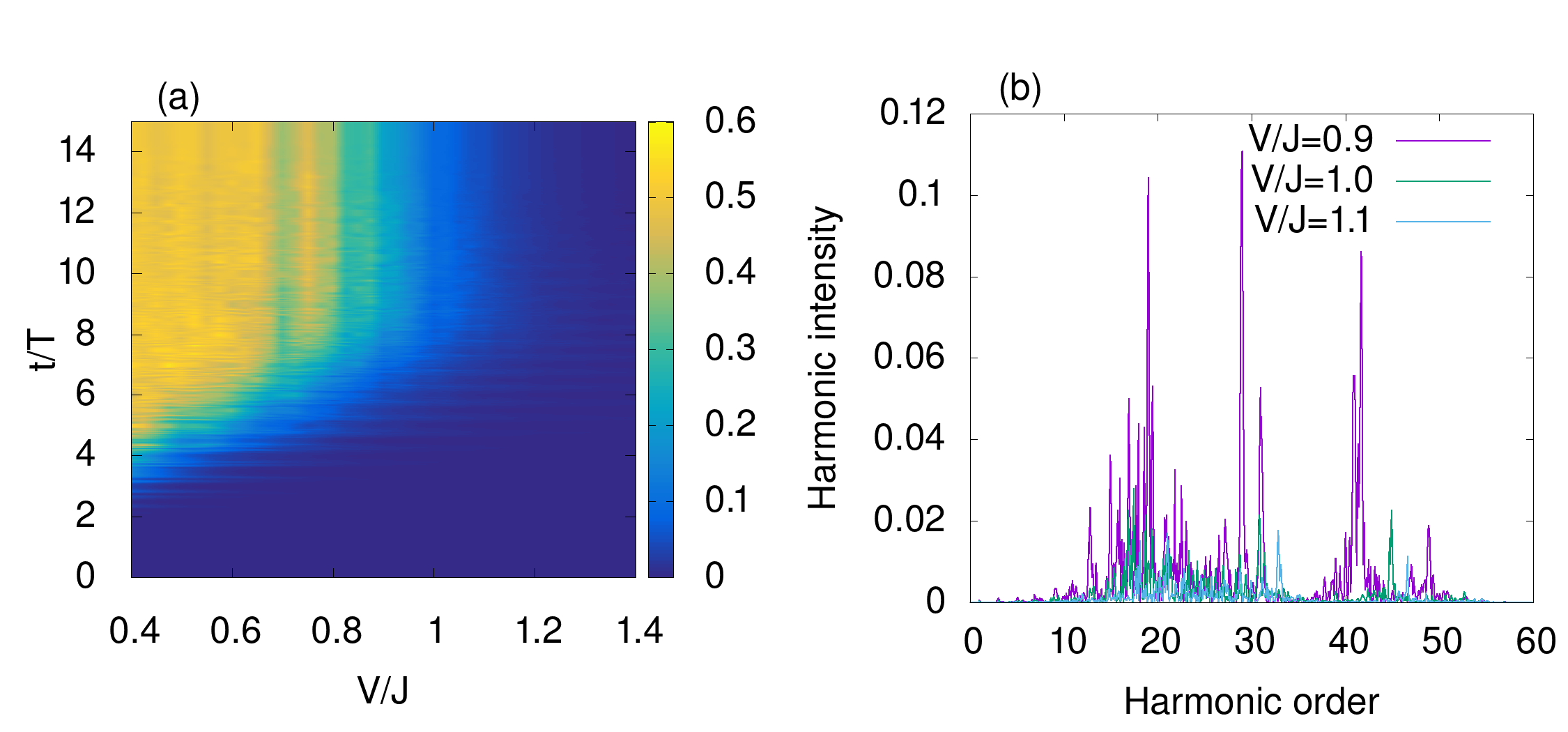}
\caption{The nonlinear dynamics of AAH model in the strong-field regime as a
function of the amplitude of the on-site potential. The fundamental
frequency is $\protect\omega =0.1J$ and the field strength is taken to be $%
E_{0}=0.4$. (a) The time evolution of the state population for the first
excited band. (b) The HHG spectra in linear scale for the values of on-site
potential at and near the criticla point.}
\end{figure}

\begin{figure}[tbp]
\includegraphics[width=0.49\textwidth]{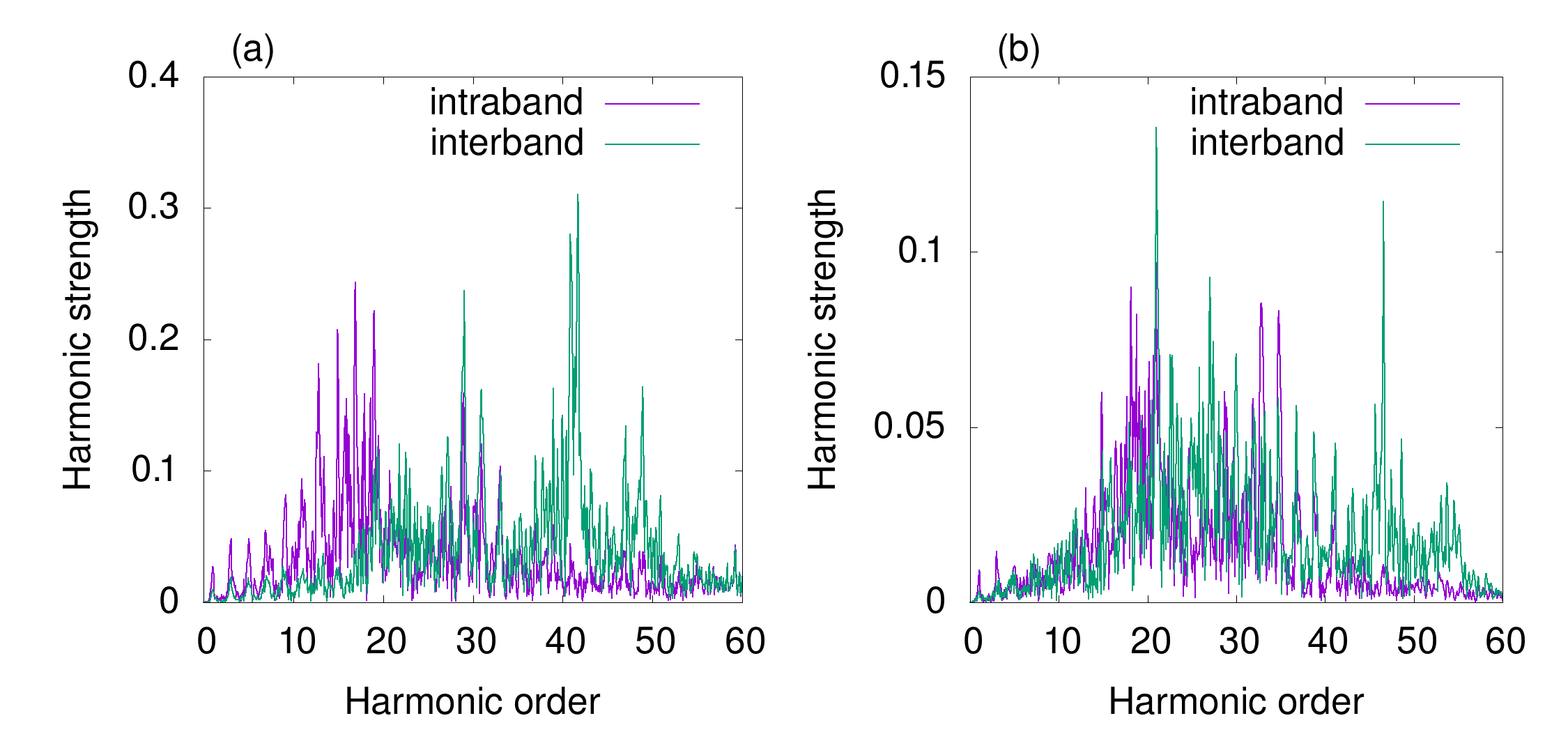}
\caption{Interband and intraband contributions to the HHG spectrum,
represented by the harmonic strengths $\left\vert a_{\mathrm{inter}}\left(
\Omega \right) \right\vert $ and $\left\vert a_{\mathrm{intra}}\left( \Omega
\right) \right\vert $, respectively. The driving field has a fundamental
frequency $\protect\omega =0.1J$ and amplitude $E_{0}=0.4$. Panel (a)
corresponds to $V/J=0.9$ and panel (b) to $V/J=1.1$.}
\end{figure}

\begin{figure*}[tbp]
\includegraphics[width=1.0\textwidth]{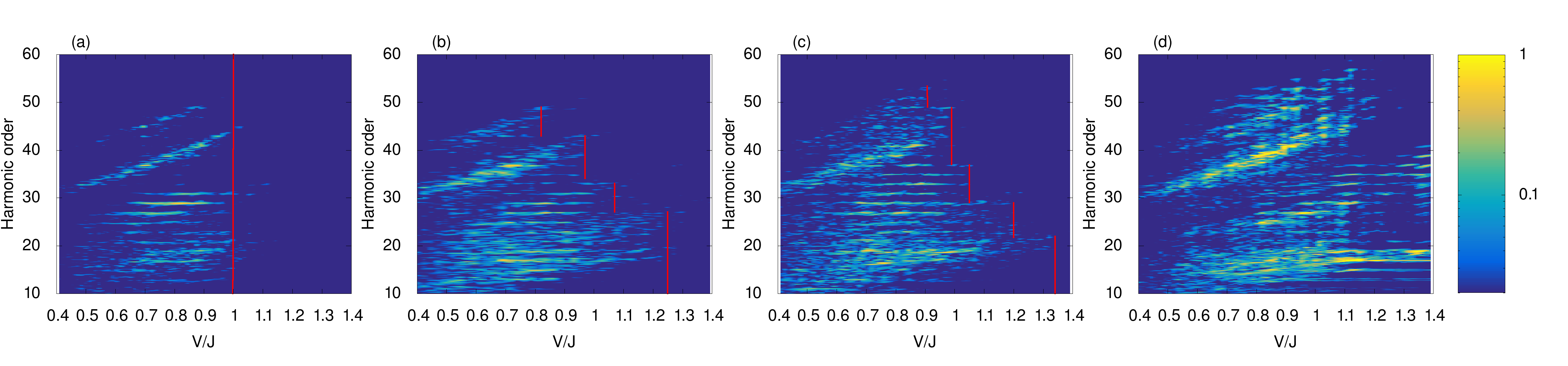}
\caption{HHG spectra in logarithmic scale for 1D quasicrystals in the
strong-field regime as a function of the amplitude of the on-site potential.
(a) AAH model; (b) BD model with $p=2.0$; (c) GPD model with $b=0.275$; (d)
AAH model with commensurate modulation $\protect\sigma =2/5$. The
fundamental frequency is $\protect\omega =0.1J$ and the field strength is $%
E_{0}=0.4$. Color scale indicates HHG intensity normalized to its maximum.
Vertical lines mark abrupt drops in the cutoff harmonics corresponding to
transitions involving states near the critical points.}
\end{figure*}

This analysis shows that by appropriately tuning the interaction parameters,
localization transitions and MEs can be effectively probed via HHG. To
validate the theoretical picture described above, we numerically simulate
the HHG response in the AAH model. Figure 5(a) displays the time evolution
of the state population in the first excited band of the AAH model, where
the band gap is approximately $2J$, as a function of the on-site potential
amplitude $V$. As expected, beyond the critical point $V=J$, the excitation
probability rapidly decreases, in agreement with the presence of
localization. For $V/J>1$, the system behaves as an ensemble of effectively
non-overlapping two-level atoms, resulting in a negligible excitation rate.
In contrast, for $V/J<1$, the states remain extended, and the excitation
probability remains high. Since excitation probability drops sharply at the
localization threshold, the HHG yield also exhibits an abrupt transition, as
illustrated in Fig. 5(b). Figure 6 shows the interband and intraband
contributions to the HHG spectra for the AAH model, based on the
decomposition given by Eqs. (\ref{dec2}) and (\ref{dec3}). As observed, the
intraband dipole acceleration dominates the low-frequency region of the
spectrum, whereas in the high-frequency range, particularly before the
critical point, the primary contribution arises from electron-hole pair
creation followed by recombination. As is seen from Figs. 5(b) and 6 the HHG
spectrum shows multiple plateau regions, each corresponding to interband
transitions involving different bands [see Fig. 3]. After crossing the
critical point, the harmonic yield decreases significantly across all
plateaus. Thus, the structure of the HHG spectrum provides direct insight
into the localization transition and the presence of MEs. For this purpose,
the laser field must be sufficiently strong to excite electrons from the
lowest energy state to the highest energy state. In the absence of
localization, the cutoff frequency at each HHG plateau is determined by the
corresponding band edge. However, the presence of a ME implies that for each
eigenenergy, there exists a critical value of $V/J$ beyond which the
eigenstates become localized. If the excitation involves transitions through
or into these localized states, a reduction in HHG yield is expected. In the
extreme case where both the initial and final states are localized, the HHG
yield can vanish entirely. To probe this further, Fig. 7 presents the HHG
spectra in the strong-field regime as a function of the on-site potential
amplitude for various 1D quasicrystal models. For comparison, we also
include the AAH model with commensurate modulation. As seen in Fig. 7(a),
for the AAH model, once $V/J>1$, the HHG yield drops sharply and eventually
vanishes. The weak residual signal near the critical point arises from
critical states at the ME, which are neither fully localized nor extended.
In contrast, for the commensurate AAH model [Fig. 7(d)], the HHG yield
remains substantial across the spectrum, even at large $V/J$ values,
reflecting the absence of localization. For the BD and GPD models [Figs.
7(b) and 7(c)], clear signatures of MEs are observed. Vertical lines in the
plots mark abrupt drops in the HHG cutoff, corresponding to transitions
involving states near the critical points identified in Figs. 3(b) and 3(c).
For the BD model, we identify multiple such drops: the first at $V/J\approx
0.8$, corresponding to transitions from the lower to upper band; the second
at $V/J\approx 0.95$, associated with the intermediate states; the third at $%
V/J\approx 1.05$, marking the localization of initial states; and the fourth
at $V/J\approx 1.25$, corresponding to the final states. Higher-band
critical points may also be detected by adjusting the filling factor to
exclude lower bands from the nonlinear dynamics. A similar structure is
observed for the GPD model, affirming the robustness of this method for
detecting MEs in quasicrystals.

To further assess the robustness of this approach for detecting MEs, we
analyze the nonlinear optical response governed by Eq.~(\ref{TDS}). First,
we diagonalize the static Hamiltonian (\ref{hc}) and compute the \textit{%
dipole mobility} for eigenstates in the lowest band. Figure~8(a) shows the
energy spectrum for $\widetilde{V}_{p}=2.0$ and $\sigma =(\sqrt{5}-1)/2$.
When $\widetilde{V}_{s}=0$, the spectrum is continuous; as $\widetilde{V}%
_{s} $ increases, it splits into minibands separated by minigaps. For $%
\widetilde{V}_{s}<0.2$, the eigenstates in the lowest band remain extended,
as indicated by maximal dipole mobility. For $\widetilde{V}_{s}>0.3$, the
eigenstates become localized and disconnected, with dipole mobility
vanishing. Thus, a series of MEs emerges within the spectrum. Importantly,
our analysis focuses only on the lowest band of the original periodic
lattice, where the secondary potential reorganizes the spectrum into
minibands.

Figure 8(b) presents the HHG spectrum in the strong-field regime as a
function of $\widetilde{V}_{s}$. The harmonic yield drops sharply around $%
\widetilde{V}_{s}\approx 0.2$ and $\widetilde{V}_{s}\approx 0.27$ and
vanishes above $\widetilde{V}_{s}>0.3$, consistent with the critical points
identified in Fig. 8(a).

\begin{figure}[tbp]
\includegraphics[width=0.5\textwidth]{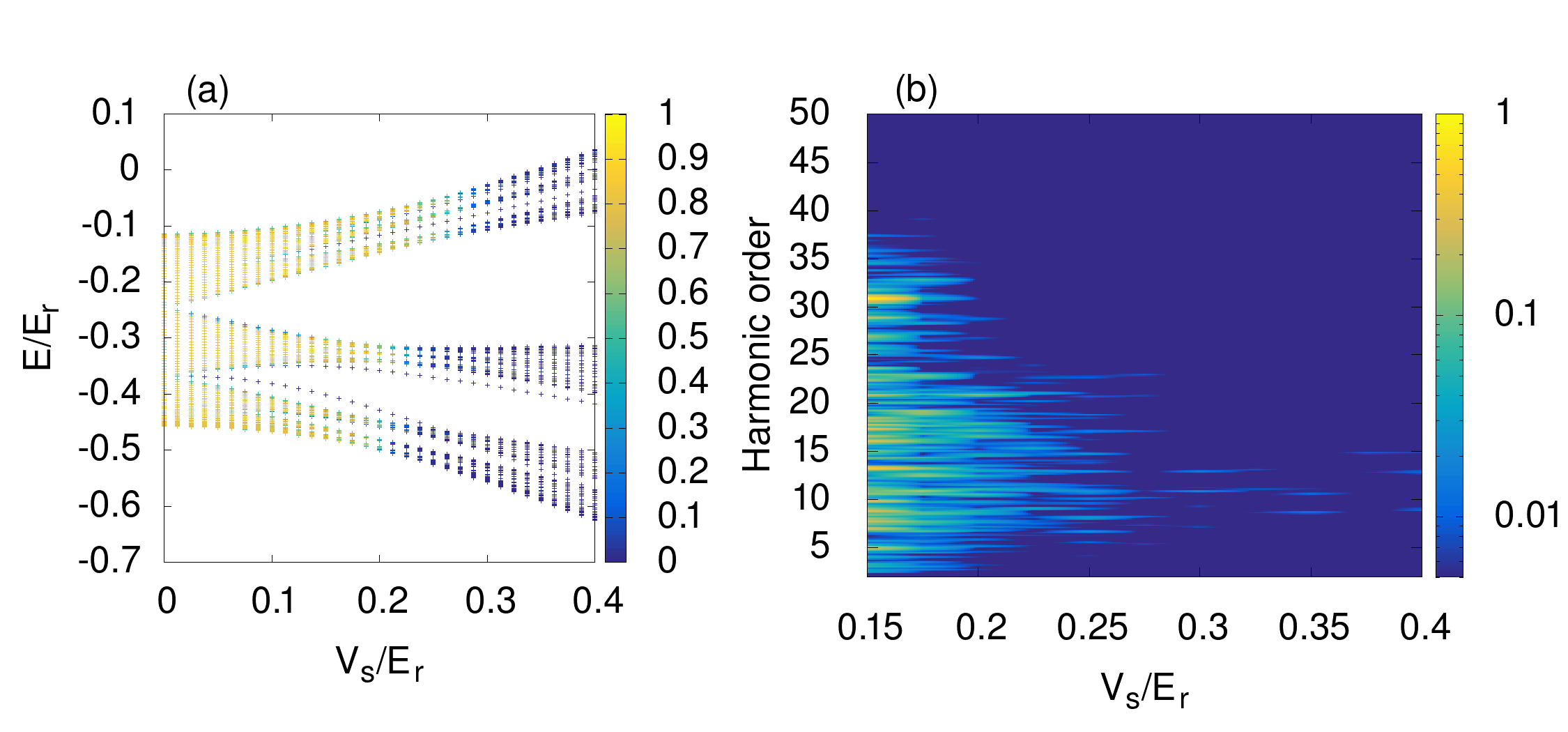}
\caption{(a) Energy eigenvalues of the 1D quasicrystal described by
Hamiltonian (\protect\ref{hc}) for $\widetilde{V}_{p}=2.0$ and $\protect%
\sigma =(\protect\sqrt{5}-1)/2$, plotted as a function of $\widetilde{V}_{s}$%
. Shading denotes dipole mobility. (b) HHG spectrum (log scale) in the
strong-field regime as a function of $\widetilde{V}_{s}$. The driving field
has frequency $\hbar \protect\omega =0.01E_{r}$, amplitude $%
elE_{0}/E_{r}=10^{-2}$, and duration $\mathcal{T}=10$ cycles. The abrupt
drops in cutoff harmonics correspond to transitions near mobility edges.}
\end{figure}

\section{Conclusion}

In conclusion, detecting MEs requires diagnostic tools beyond conventional
linear-response techniques, which pose both theoretical and experimental
challenges. Our results demonstrate that HHG provides a sensitive, nonlinear
optical probe of localization transitions and MEs in quasiperiodic systems.
By introducing the concept of dipole mobility, we offer a simple and
intuitive framework that connects intraband dynamics with HHG yields across
different localization regimes. This approach is largely model-independent
and can be extended to more complex and realistic systems, including
three-dimensional structures. These results open promising new possibilities
for probing quasiperiodic and disordered systems through nonlinear
spectroscopy, contributing to the expanding field of extreme solid-state
photonics. Furthermore, we emphasize that, in quasiperiodic systems, the
absence of translational invariance means that Bloch momentum is not a good
quantum number. This fundamental distinction from periodic crystals
underlies the breakdown of Bloch band theory and necessitates alternative
approaches--such as grouping eigenstates into contiguous energy intervals
separated by large gaps--when analyzing optical and nonlinear responses.
Recognizing this distinction is essential for correctly interpreting
``interband" and ``intraband" pathways in HHG for such systems. Although the
present study is restricted to one-dimensional models, an important
direction for future research is to generalize this framework to
higher-dimensional systems and incorporate electron-electron interactions.

\begin{acknowledgments}
The work was supported by the Science Committee of Republic of
Armenia, project No. 21AG-1C014.
\end{acknowledgments}

\appendix

\section{Multiphoton and tunneling excitation regimes}

In contrast to periodic systems, quasiperiodic lattices lack translational
symmetry, and Bloch momentum is therefore not a good quantum number. As a
result, the concepts of a crystal Brillouin zone and continuous band
dispersion do not apply in the strict sense. Instead, the single-particle
spectrum exhibits a fractal hierarchy of subbands, with gaps that depend
sensitively on the incommensurate modulation parameter. Throughout this
work, the term ``band" is used to denote a contiguous energy interval
separated from others by gaps significantly larger than the driving photon
energy. Such grouping provides a useful working definition for analyzing
optical and nonlinear responses, even though the underlying states do not
possess Bloch-like momentum labeling.

In this Appendix, we examine excitation regimes between two such energy
intervals (``bands") that exhibit distinct localization properties of their
single-particle wavefunctions. Our focus is on the case of a low-frequency
driving laser field, where both interband and intraband pathways can play a
role in the resulting high-harmonic response. This analysis provides the
theoretical underpinning for the physical interpretation of the HHG cutoff
behavior discussed in the main text.

\subsection{Multiphoton perturbative excitation regime}

When the states within both bands are fully localized, i.e., $D_{\mu }=0$
the system behaves as a disjoint set of two-level atoms (see Fig. 4(a) in
the main text). In this regime, excitation occurs via the multiphoton
perturbative process. From Eq. (6) in the main text, since $D_{\mu }=0$, for
a particular pair of states (e.g., $\left\vert 1\right\rangle $ and $%
\left\vert 2\right\rangle $), we obtain the following equations:

\begin{equation*}
i\frac{\partial \varrho _{12}}{\partial t}=\left( \varepsilon
_{1}-\varepsilon _{2}\right) \varrho _{12}+E\left( t\right) d_{12}\left(
\varrho _{22}-\varrho _{11}\right) ,
\end{equation*}%
\begin{equation*}
i\frac{\partial \varrho _{22}}{\partial t}=E\left( t\right) \left( \varrho
_{12}d_{21}-\varrho _{21}d_{12}\right) ,
\end{equation*}%
\begin{equation}
\varrho _{11}=1-\varrho _{22}.  \label{1s}
\end{equation}%
This system of equations is equivalent to the evolution equations for the
state amplitudes $a_{1}$ and $a_{2}$:%
\begin{equation}
i\frac{d}{dt}\left[ 
\begin{array}{c}
a_{1} \\ 
a_{2}%
\end{array}%
\right] =\left[ 
\begin{array}{cc}
\varepsilon _{1} & E\left( t\right) d_{12} \\ 
E\left( t\right) d_{21} & \varepsilon _{2}%
\end{array}%
\right] \left[ 
\begin{array}{c}
a_{1} \\ 
a_{2}%
\end{array}%
\right] ,  \label{2s}
\end{equation}%
with $\varrho _{12}=a_{1}a_{2}^{\ast }$, $\varrho _{11}=\left\vert
a_{1}\right\vert ^{2}$, and $\varrho _{22}=\left\vert a_{2}\right\vert ^{2}$%
. \ 

This system describes Rabi oscillations when the wave is in one-photon
resonance with the two-level system, i.e., $\omega =\Delta \equiv
\varepsilon _{2}-\varepsilon _{1}$ and $\left\vert E_{0}d_{12}\right\vert
<<\Delta $. When $\omega <<\Delta $, one can consider multiphoton Rabi
oscillations, treating the system by using Floquet theory \cite%
{shirley1965solution} or the Wentzel-Kramers-Brillouin representation in the
complex time plane \cite{duvall1988nonperturbative}. Both approaches yield
similar results. Assuming the system initially occupies state $\left\vert
1\right\rangle $, for an $n$-photon resonant excitation $n=\Delta /\omega $
, the population is given by: 
\begin{equation}
\varrho _{22}\simeq \frac{\Omega _{n}^{2}}{\frac{\delta _{n}^{2}}{4}+\Omega
_{n}^{2}}\sin ^{2}\sqrt{\frac{\delta _{n}^{2}}{4}+\Omega _{n}^{2}}t,
\label{3s}
\end{equation}%
where 
\begin{equation}
\Omega _{n}=\frac{\omega }{\pi }\left( \frac{E_{0}\left\vert
d_{12}\right\vert }{2\Delta }e_{0}\right) ^{\frac{\Delta }{\omega }}
\label{4s}
\end{equation}%
is the $n$-photon resonant Rabi frequency, $e_{0}$ is the base of the
natural logarithm, and 
\begin{equation*}
\delta _{n}=\Delta -n\omega +\frac{E_{0}^{2}\left\vert d_{12}\right\vert ^{2}%
}{\Delta }
\end{equation*}%
is the resonant detuning. For the system under consideration, we obtain \ $%
\Omega _{n}/\omega <10^{-17}$, leading to a negligible excitation
probability.

\subsection{Tunneling Excitation Regime}

Now, we consider the tunneling excitation regime (see Fig. 4(c) in the main
text), where the states within one band are fully delocalized ($D_{\mu
}\simeq 1$). To determine the HHG probability in the presence of the field $%
E\left( t\right) $, we start from the wave equation:%
\begin{equation}
i\frac{\partial \left\vert \Psi \right\rangle }{\partial t}=\left( \frac{1}{2%
}\widehat{p}^{2}+U\left( x\right) -xE\left( t\right) \right) \left\vert \Psi
\right\rangle .  \label{weq}
\end{equation}%
Here we employ local atomic units. The localized state wavefunction is
modeled as: 
\begin{equation*}
\left\vert l\right\rangle =\frac{1}{\pi ^{1/4}\Gamma ^{1/2}}e^{-\frac{x^{2}}{%
2\Gamma ^{2}}},
\end{equation*}%
where $\Gamma $ defines the localization width. The localization is enforced
by an assumed short-range potential $U\left( x\right) $. For the extended
state, we consider a de Broglie wave: 
\begin{equation*}
\left\vert p\right\rangle =\frac{1}{\left( 2\pi L\right) ^{1/2}}e^{ipx},
\end{equation*}%
where $p$ is the momentum, which serves as the quantum number defining the
continuum. An alternative approach using standing waves is also possible,
but it leads to the same conclusion. For the localized state, 
\begin{equation*}
\left( -\frac{1}{2}\frac{d^{2}}{dx^{2}}+U\left( x\right) \right) \left\vert
l\right\rangle =\varepsilon \left\vert l\right\rangle ,
\end{equation*}%
where $\varepsilon $ is the eigenenergy. For extended states, the potential
is neglected, leading to: 
\begin{equation*}
\left( -\frac{1}{2}\frac{d^{2}}{dx^{2}}\right) \left\vert p\right\rangle =%
\frac{p^{2}}{2}\left\vert p\right\rangle .
\end{equation*}%
Following the ansatz in \cite{lewenstein1994theory}, we expand the
time-dependent wavefunction as:%
\begin{equation}
\left\vert \Psi \right\rangle =\left( C_{0}\left( t\right) \left\vert
l\right\rangle +\int dpC\left( p\mathbf{,}t\right) \left\vert p\right\rangle
\right) e^{-i\varepsilon t}.  \label{1}
\end{equation}%
Neglecting the depletion of the initial state, $C_{0}\left( t\right) \simeq
1,$ and from Eq. (\ref{weq}) we derive the probability amplitude equation
for $C\left( p\mathbf{,}t\right) $:%
\begin{equation*}
\frac{\partial C\left( p\mathbf{,}t\right) }{\partial t}+E\left( t\right) 
\frac{\partial C\left( p\mathbf{,}t\right) }{\partial p}
\end{equation*}%
\begin{equation}
+i\left( \frac{p^{2}}{2}-\varepsilon \right) C\left( p\mathbf{,}t\right) =i%
\mathcal{D}\left( p\right) E\left( t\right) ,  \label{CC}
\end{equation}%
where $\mathcal{D}\left( p\right) =\left\langle l\right\vert x\left\vert
p\right\rangle $ is the dipole matrix element for the localized-extended
state transition, given by: 
\begin{equation*}
\mathcal{D}\left( p\right) =\frac{\Gamma ^{5/2}}{L^{1/2}\pi ^{1/4}}ipe^{-%
\frac{p^{2}\Gamma ^{2}}{2}}.
\end{equation*}%
Note that the second term in Eq. (\ref{CC}) is just conditioned by the
dipole mobility ($D_{\mu }\simeq 1$). Then, from Eq. (\ref{CC}) for the
amplitude $C\left( p\mathbf{,}t\right) $ we obtain:%
\begin{equation*}
C\left( p\mathbf{,}t\right) =i\int_{0}^{t}dt^{\prime }\mathcal{D}\left(
p+\left( A\left( t\right) \mathcal{-}A\left( t^{\prime }\right) \right)
\right) E\left( t^{\prime }\right)
\end{equation*}%
\begin{equation}
\times \exp \left\{ -i\int_{t^{\prime }}^{t}\left[ \frac{1}{2}\left( p%
\mathbf{+}A\left( t\right) \mathbf{-}A\left( t^{\prime \prime }\right)
\right) ^{2}-\varepsilon \right] dt^{\prime \prime }\right\} ,  \label{Cp}
\end{equation}%
where $A\left( t\right) =-\frac{E_{0}}{\omega }\sin \omega t$ is the vector
potential of the wave field. For the harmonic radiation one needs the mean
value of the $x$: $x\left( t\right) =\left\langle \Psi \right\vert
x\left\vert \Psi \right\rangle $. Using Eqs. (\ref{1}) and (\ref{Cp}), we
obtain: 
\begin{equation*}
x\left( t\right) =i\int dp\int_{0}^{t}dt_{1}\mathcal{D}\left( p\mathbf{-}%
A\left( t_{1}\right) \right) E\left( t_{1}\right)
\end{equation*}%
\begin{equation*}
\times \mathcal{D}^{\ast }\left( p-A\left( t\right) \right)
\end{equation*}%
\begin{equation}
\times \exp \left\{ -iS\left( p,t,t_{1}\right) \right\} +\mathrm{c.c.},
\label{xt}
\end{equation}%
where 
\begin{equation}
S(p\mathbf{,}t,t_{1})=\int_{t_{1}}^{t}\frac{1}{2}\left( \left( p\mathbf{-}%
A\left( t_{2}\right) \right) ^{2}-\varepsilon \right) dt_{2}.  \label{action}
\end{equation}

As in the atomic case \cite{lewenstein1994theory}, the HHG rate is mainly
determined by the exponential in the integrand of Eq. (\ref{xt}). The
integral over the intermediate momentum $p$ and time $t_{1}$ can be
calculated using the saddle-point method. The saddle momentum\ is determined
by the equation:%
\begin{equation}
\frac{\partial S(p\mathbf{,}t,t_{1})}{\partial p}=0,  \label{sad1}
\end{equation}%
hence the saddle momentum ($p_{s}$) is given by the solution of the
equation: 
\begin{equation}
\int_{t_{1}}^{t}\left( p_{s}\mathbf{-}A\left( t_{2}\right) \right) dt_{2}=0.
\label{sad_r}
\end{equation}%
Integrating in Eq. (\ref{xt}) over $p$, for the mean dipole moment we obtain:%
\begin{equation*}
x(t)=\left( 2\pi \right) ^{1/2}\sqrt{i}\int_{0}^{t}dt_{1}\frac{e^{-i\left(
S(p_{s}\mathbf{,}t,t_{1})-\varepsilon \left( t-t_{1}\right) \right) }}{\sqrt{%
\left\vert t-t_{1}\right\vert }}E\left( t_{1}\right)
\end{equation*}%
\begin{equation}
\times \mathcal{D}\left( p_{s}\mathcal{-}A\left( t_{1}\right) \right) 
\mathcal{D}^{\ast }\left( p_{s}-A\left( t\right) \right) +\mathrm{c.c. .}
\label{xt2}
\end{equation}%
The complex saddle times $t_{s}$ are the solutions of the following
equation: 
\begin{equation}
\frac{\partial S(p_{s}\mathbf{,}t,t_{1})}{\partial t_{1}}+\varepsilon =0,
\label{sad2}
\end{equation}%
which may be expressed by the transcendental equation 
\begin{equation}
\frac{1}{2}\left( p_{s}\mathbf{-}A\left( t_{s}\right) \right)
^{2}-\varepsilon =0.  \label{sad22}
\end{equation}%
Then expressing the saddle time as $t_{s}=t_{b}+i\delta $, with $\omega
\delta <<1$, one can obtain the saddle momentum: 
\begin{equation}
p_{s}=A\left( t_{b}\right) ,  \label{ps}
\end{equation}%
and the imaginary part of the saddle time:%
\begin{equation}
\delta =\frac{\sqrt{2\left\vert \varepsilon \right\vert }}{\left\vert
E\left( t_{b}\right) \right\vert },  \label{imag}
\end{equation}%
where $E\left( t_{b}\right) =E_{0}\cos \omega t_{b}$. Taking into account
Eq. (\ref{ps}), from Eq. (\ref{sad_r}) for the real part of the saddle time
we obtain: 
\begin{equation}
\int_{t_{b}}^{t}\left( \sin \omega t_{2}-\sin \omega t_{b}\right) dt_{2}=0.
\label{Ret}
\end{equation}

As usual, $t_{b}$ is interpreted as the birth time of the photoelectron
which returns at the moment $t$ to the core and generates harmonic
radiation. Thus, we obtain the ultimate formula for the dipole moment:%
\begin{equation}
x(t)=\sum\limits_{t_{b}}C_{\mathrm{ion}}\left( t_{b}\right) C_{\mathrm{pr}%
}\left( t,t_{b}\right) C_{\mathrm{rec}}\left( t,t_{b}\right) +\mathrm{c.c.}.
\label{maind}
\end{equation}

Formula (\ref{maind}) is analogous to the nonrelativistic formula for the
dipole moment in the three-step model \cite{ivanov1996coulomb}. Here the
summation is carried out over the solutions of Eq. (\ref{Ret}). The
tunneling ionization amplitude $C_{\mathrm{ion}}\left( t_{b}\right) $ is: 
\begin{equation}
C_{\mathrm{ion}}\left( t_{b}\right) =\frac{\Gamma ^{5/2}}{L^{1/2}\pi ^{1/4}}%
\sqrt{2\left\vert \varepsilon \right\vert }e^{\left\vert \varepsilon
\right\vert \Gamma ^{2}}e^{-\frac{\left( 2\left\vert \varepsilon \right\vert
\right) ^{3/2}}{3\left\vert E\left( t_{b}\right) \right\vert }}.
\label{Cion}
\end{equation}%
\ The propagation amplitude is given by the expression: 
\begin{equation}
C_{\mathrm{pr}}\left( t,t_{b}\right) =\left( 2\pi i\right) ^{1/2}\frac{\exp
\left\{ -iS\left( p_{s}\mathbf{,}t,t_{b}\right) +i\varepsilon \left(
t-t_{b}\right) \right\} }{\sqrt{\left\vert t-t_{1}\right\vert }},
\label{Cpr}
\end{equation}%
and the recombination amplitude is: 
\begin{equation}
C_{\mathrm{rec}}\left( t,t_{b}\right) =\frac{\Gamma ^{5/2}}{L^{1/2}\pi ^{1/4}%
}i\left( p_{s}-A\left( t\right) \right) e^{-\frac{\left( p_{s}-A\left(
t\right) \right) ^{2}\Gamma ^{2}}{2}}.  \label{Crec}
\end{equation}

As is seen from Eqs. (\ref{Cion}), (\ref{Cpr}), and (\ref{Crec}), the
localization effects are considerable for ionization and recombination.
Returning to our model considered in the main text, we can estimate the
local atomic field strength, as $E_{a}/E_{0}\simeq \Delta /d_{tr}E_{0}\simeq
\allowbreak 20.0$. For the tunneling time we have $\omega \delta =0.125$ and
for the exponential factor we have $\exp \left\{ -\left( 2\left\vert
\varepsilon \right\vert \right) ^{3/2}/3\left\vert E\left( t_{b}\right)
\right\vert \right\} \sim 10^{-3}\ $which provides reasonable values for
excitation compared with the disjoint set of two level atoms.

Let us now consider the tunneling excitation regime when the states within
both bands are fully delocalized, as shown in Fig. 4(b). This situation
resembles the excitation process in periodic crystals \cite%
{vampa2014theoretical}, where interband tunneling leads to the formation of
electron-hole pairs. The saddle-point analysis for such a scenario requires
explicit consideration of band dispersion and Bloch-like wavefunctions, as
detailed in \cite{vampa2014theoretical}. Nonetheless, the key insight from
our analysis remains valid. In this context, the final energy in the action
integral (Eq. (\ref{action})) corresponds to the sum of the electron and
hole energies: 
\begin{equation}
S(p\mathbf{,}t,t_{1})=\int_{t_{1}}^{t}\varepsilon _{eh}\left( p\mathbf{-}%
A\left( t_{2}\right) \right) ^{2}dt_{2}.  \label{eha}
\end{equation}%
The subsequent analysis follows the same logic as in the atomic-like
scenario (Fig. 4(c)), with the dispersion relation near the band edge
approximated as: 
\begin{equation*}
\varepsilon _{eh}\left( p\right) =\Delta +\frac{p^{2}}{2}.
\end{equation*}%
Consequently, the action retains the same exponential structure as in the
atomic tunneling case, but with the binding energy $\left\vert \varepsilon
\right\vert $ replaced by the band gap $\Delta $. While the pre-exponential
factor differs due to the nature of the band structure, the qualitative
behavior of the excitation amplitude is preserved. Importantly, the number
of available excitation and recombination channels is larger in this case,
enhancing the overall HHG yield relative to the atomic-like scenario in Fig.
4(c).

\bibliographystyle{apsrev4-2}
\bibliography{bibliography}

\end{document}